\documentclass[a4paper,fleqn]{cas-sc}

\usepackage[authoryear,longnamesfirst]{natbib}

\usepackage{hyperref}
\usepackage{multirow}
\usepackage{graphicx} 
\usepackage{subfigure}
\usepackage{paralist}
\usepackage{xcolor}

\def\tsc#1{\csdef{#1}{\textsc{\lowercase{#1}}\xspace}}
\tsc{WGM}
\tsc{QE}
\tsc{EP}
\tsc{PMS}
\tsc{BEC}
\tsc{DE}

\begin{document}
\let\WriteBookmarks\relax
\def\floatpagepagefraction{1}
\def\textpagefraction{.001}

\shorttitle{Characterizing Multi-domain False News and Underlying User Effects on Chinese Weibo}

\shortauthors{Sheng et~al.}

\title [mode = title]{Characterizing Multi-domain False News and Underlying User Effects on Chinese Weibo}           

%
\author[ict,ucas]{Qiang Sheng}
[orcid=0000-0002-2481-5023]
\credit{Conceptualization, Methodology, Software, Validation, Formal analysis, Investigation, Data curation, Writing -- original draft} 
\ead{shengqiang18z@ict.ac.cn}

\author[ict,ucas]{Juan Cao}
\cormark[1]
\cortext[cor1]{Corresponding author.}
\credit{Conceptualization, Methodology, Resources, Supervision, Funding acquisition} 
\ead{caojuan@ict.ac.cn}

\author[asu-bernard]{H. Russell Bernard}
\credit{Conceptualization, Writing -- review \& editing} 
\ead{asuruss@asu.edu}

\author[iit]{Kai Shu}
\credit{Conceptualization, Writing -- review \& editing} 
\ead{kshu@iit.edu}

\author[ict]{Jintao Li}
\credit{Project administration}
\ead{jtli@ict.ac.cn}

\author[asu-liu]{Huan Liu}
\credit{Conceptualization, Writing -- review \& editing}
\ead{huanliu@asu.edu}

\address[ict]{Key Lab of Intelligent Information Processing, Institute of Computing Technology, Chinese Academy of Sciences, Beijing, China}
\address[ucas]{University of Chinese Academy of Sciences, Beijing, China}
\address[asu-bernard]{Institute for Social Science Research, Arizona State University, Tempe, AZ, USA}
\address[iit]{Department of Computer Science, Illinois Institute of Technology, Chicago, IL, USA}
\address[asu-liu]{Computer Science and Engineering, Arizona State University, Tempe, AZ, USA}

\begin{abstract}
False news that spreads on social media has proliferated over the past years and has led to multi-aspect threats in the real world.
While there are studies of false news on specific domains (like politics or health care), little work is found comparing false news across domains.
In this article, we investigate false news across nine domains on Weibo, the largest Twitter-like social media platform in China, from 2009 to 2019. The newly collected data comprise 44,728 posts in the nine domains, published by 40,215 users, and reposted over 3.4 million times.
Based on the distributions and spreads of the multi-domain dataset, we observe that false news in domains that are close to daily life like health and medicine generated more posts but diffused less effectively than those in other domains like politics,
and that political false news had the most effective capacity for diffusion.
The widely diffused false news posts on Weibo were associated strongly with certain types of users -- by gender, age, etc. Further, these posts provoked strong emotions in the reposts and diffused further with the active engagement of false-news starters.
Our findings have the potential to help design false news detection systems in suspicious news discovery, veracity prediction, and display and explanation.
The comparison of the findings on Weibo with those of existing work demonstrates nuanced patterns, suggesting the need for more research on data from diverse platforms, countries, or languages to tackle the global issue of false news.
The code and new anonymized dataset are available at \url{https://github.com/ICTMCG/Characterizing-Weibo-Multi-Domain-False-News}.
\end{abstract}


\maketitle

\section{Introduction} \label{intro}
Social media are now long established as a daily source of news in many countries around the world, Western or Eastern, developed or developing~\citep{pew-news-use, china-news-use}. These platforms facilitate equally the distribution of both reliable news as well as false news (including fake news). The problem with false and fake news on social media is widely documented. These include threats to the economy~\citep{obama}, to social order~\citep{salt, shuanghuanglian}, to politics~\citep{pizzagate}, and to physical security~\citep{India-mob, Bangladesh-lynchings}. Efforts to mitigate the spread of false news by researchers in social, political, and computer science include exploring the characteristics of false news~\citep{science18, science19, user-profile, pnas16}, detecting false news using machine learning techniques~\citep{castillo11, majing16, multimodal, defend}, and developing the automatic detection and verification system~\citep{zhouxing2015, credeye, defend-system}. Among these efforts, empirical studies for characterizing false news are fundamental to both reveal the phenomenon and to guide the design of detection methods.

Existing empirical studies have examined false news either in general~\citep{science18}, or in a specific domain, such as politics~\citep{science19, sa19, user-profile}, science~\citep{pnas16}, health~\citep{fakecures}, and entertainment~\citep{user-profile}.
While comparisons of false news across diverse domains are rare and limited \citep{mdfend, embracing}, one of the findings in ~\citep{science18} suggests the need for domain-level spread analysis: On Twitter,\footnote{\url{https://twitter.com/}} for example, political false news had more effective capacity for diffusion than any other on Twitter (it traveled farther and reached more people than non-political false news). But so far, it remains under-explored how false news in other domains spread, which is important for highlighting and positioning the influence of false news in different domains and guiding the design of detection systems.
In this article, we use a new dataset of 44,728 false social media posts from Weibo,\footnote{\url{https://weibo.com/}} the largest Twitter-like social media platform in China, to investigate the capacity for diffusion of false and fake news posts in nine domains: (1) Society \& Life, (2) Disasters \& Accidents, (3) Health \& Medicine, (4) Education \& Examinations, (5) Culture \& Sports \& Entertainment, (6) Science \& Technology, (7) Finance \& Business, (8) Politics, and (9) Military. We then explore how user characteristics (such as gender, age, and account type) are related to the spread process and how user emotions and behaviors affect the spread of these false posts. Our contributions are as follows:

\begin{itemize}
	\item \textbf{Capacity for diffusion.} We find that false news on life-unrelated domains generated fewer stories but diffused better than those on life-related domains. Of the nine domains, political false news had the most effective capacity for diffusion.
	\item \textbf{User effects.} We characterize the user effects of the widely diffused false stories: They engaged more males, older users, or verified users. Further, they provoked strong emotions in the reposts and diffused along with false-news starters' active engagements.
	\item \textbf{Methodology.} This work introduces multi-domain analysis, a new perspective to understand the phenomenon of false news. We design the rules to rank the domain-level capacity for diffusion and then observe the user effects by statistical, linguistic, and semantical measurements.
	\item \textbf{Data.} We collect a multi-domain false news dataset from Chinese Weibo, which contains 44,728 false stories of nine domains from 2009 to 2019. To the best of our knowledge, the dataset has the largest amount and is for the longest period for false news research on Chinese social media.
\end{itemize}

\section{Related Work and Research Questions} \label{related-work}

\subsection{Definition of False News}
The recent attention to this field is largely due to so-called fake news going viral during the 2016 U.S. presidential election~\citep{lie-of-2016-fake-news}. The term ``fake news'' and other related concepts including \textit{rumor}, \textit{misinformation}, \textit{disinformation}, and \textit{false news} are used in published studies interchangeably to describe the social media posts we are working with. We rely here on the definitions in \citep{science18} and \citep{zhouxinyi-tutorial} and define \textit{false news} as any story or claim with a false assertion with unknown intention. A \textit{rumor} refers to an unverified and instrumentally relevant statement of information spread among people~\citep{shukai-survey} and is not germane to our research interest on verified inaccurate information. \textit{Fake news} and \textit{disinformation} refer to intentionally false information~\citep{jep17, science18, shukai-survey, zhouxinyi-tutorial} where the true intention of creators is hard to know in ex post collection. \textit{Misinformation} is a broad term, since it includes \textit{any} inaccurate posts~\citep{mohseni2019open}, which, therefore, our method for data collection could not cover. \textit{False news} includes disinformation as well as well-intentioned, but untrue news stories, which is proper in our work.

\subsection{False News Characterization}
The openness and freedom of access on social media provides researchers with observable, naturally occurring, large-scale data without conducting individual-level interviews or in-lab experiment. Researchers in this arena first collect news posts from the official data interfaces or webpage parsers of social media platforms and label the posts as true or false according to the rating from reliable fact-checking organizations (e.g., Snopes\footnote{\url{https://snopes.com/}} in the United States and Jiaozhen\footnote{\url{https://fact.qq.com/} and \url{https://new.qq.com/omn/author/5107513}} in China). Then the data along with social contexts are analyzed using statistical methods for new findings. According to~\citep{zhang-survey}, four major components are involved in false news: creator/spreader, target victims, news content, and social context. As news content and social context are often closely related to fake news detection, we will detail these components along with the detection methods. Here, we introduce the researches on the spread of false news and the involved users.

Research on the spread of false news exhibits false news sharing and further influences, especially in important events such as elections and pandemics. \citet{jep17} found that pro-Trump fake stories were more widely shared on Facebook than pro-Clinton ones before the 2016 U.S. presidential election. \citet{portuguese} found that fake news is more likely to be shared but got fewer reactions than real news before the 2019 Portuguese election. The network analysis by \citet{memon-covid} suggested that misinformation on COVID-19 spreads in denser and more organized communities than true information. Instead of focusing on a specific event, our first research question is inspired by~\citep{science18}, which measured the spread of false news in diverse domains on Twitter from 2006 to 2017 and found that political false news spread faster and deeper than non-political false news. We extend this research to comparing across the nine domains noted above: \textbf{RQ1) Are there differences in the capacity for diffusion of false news in different domains?} We measured the capacity for diffusion of all domains and showed which domain(s) of false news warrant monitoring for real-time detection and mitigation.

The creators and spreaders play important roles in false news spread. \citet{user-profile3} found that verified accounts with a large number of friends posted false rumors with a small probability. \citet{rampersad2020fake} found that age has a strong influence on the acceptance of fake news in Saudi Arabia. \citet{science19} analyzed respondents' Facebook\footnote{\url{https://www.facebook.com/}} sharing history during the 2016 U.S. presidential campaign and found strong user effects: Older users shared more fake posts and the super-sharers of fake news sources were disproportionately female and unverified. We are interested in how the user effects will be in an observation across domains: \textbf{RQ2) How are the demographic factors, like gender, age, and verification status, related to the spread of false news in different domains?} Extending this to nine domains can help us understand, at a more granular level, the kind of users who tend to engage in false news and could advance mitigation by guiding the initial filtering of users susceptible to false news in specific domains.

\subsection{False News Detection}
Though manual fact-checking by journalists or independent organizations is still the most prevalent method to debunk false news, automatic detection of false news (i.e., predicting the veracity of a given post) by machine-learning techniques has become a promising direction due to its expected high computing efficiency and low labor costs. The methods can be divided into two genres: knowledge-based and appearance-based.~\citep{sheng-cikm}

Knowledge-based methods for predicting veracity start by collecting evidence and then applying reasoning, but the sources of evidence are diverse. Comment-based methods employed crowd wisdom for prediction \citep{CSI, defend, QSAN}. For those that have been previously fact-checked claims, debunking articles are used for matching with news posts \citep{shaar, nguyen-vo, sheng, ecir}. The scope of evidence is broadened to evidential web articles~\citep{declare,wulianwei-ijcai,wulianwei-tkde} or contemporary mainstream news~\citep{sheng-acl22}. Instead of obtaining in-the-wild knowledge, recent works leveraged entity background information obtained from knowledge graphs \citep{health-misinformation, MKEMN, comparenet}. For multi-modal scenarios, entity knowledge is important to bridge the text-image semantics \citep{junxiaoxue-ipm, qipeng-crad, qipeng-mm, lipeiguang-tmm}. These methods could provide accurate and explainable evidence, but have the issue of source credibility and scalability.

Instead of focusing on what the publisher says, appearance-based methods focus more on how false news \textit{looks} different from true news. The differences are captured from multiple perspectives such as content styles~\citep{style-aaai, sigir22}, emotional signals~\citep{dual-emotion}, user credibility~\citep{user-profile2}, and audiences' behaviors (e.g., \textit{like}, repost, and make comments)~\citep{defend}.
\citet{user-profile2} utilized user profiles that contain metadata on personal pages and inferred demographic features to detect fake news.
Our findings on \textbf{RQ2} can clarify the effects of some key demographic features across the domains and indicate the application areas of such methods. 
In terms of emotional signals, several independent works~\citep{science18, ajao19, dual-emotion, moral-emotion} found the statistically significant different between fake and real news. For instance, \citet{science18} calculated emotion vectors for reply tweets based on an emotion word lexicon and found that false rumors inspired replies expressing greater surprise and disgust. \citet{moral-emotion} found that COVID-19 misinformation is more likely to go viral than truthful information, especially the original posts expressing contempt, anger, and disgust. \citet{dual-emotion} performed a significant test between real and fake news on Chinese Weibo using a diverse emotion feature set of the contents and comments and showed that the emotion signals statistically correlate to news veracity. The emotional features were then used to improve the performance of text-based fake news detectors~\citep{dual-emotion, sheng-cikm}. Our third research question focuses on the role of these emotional signals: \textbf{RQ3) How are the emotional signals related to the spread of false news among the domains?} Our findings would provide a new understanding of the effects of emotion signal at the domain-level.
Behavior-based methods modeled the propagation network where nodes are connected based on the user behaviors and captured the unique diffusion patterns of false news as predictors~\citep{majing16, majing17, diffusion-alone, GCAN, CED, retweet-dynamics}. In this article, we research the only behaviors a user can employ to enlarge the size of cascades, i.e., reposting. The fourth research question, then, is: \textbf{RQ4) What did the engaged users do that promoted the spread of false news in each domain?} We expect to characterize reposting behaviors in domains where posts generally diffuse widely and find the key users that promote the diffusion.

Overall, our work on \textbf{RQ1} measures the spread of false news in different domains, and the rest of the research questions (\textbf{RQ2}, \textbf{RQ3}, and \textbf{RQ4}) expose the relationship between the user engagements and the spread of various kinds of false news.
\section{Data} \label{data}
 
\subsection{Platform Selection}
We use data from Weibo because of its richness, its comparability with Twitter, and its accessibility. Weibo has been providing microblogging service in China since August 2009~\citep{wiki-weibo}, and is now the largest microblogging platform in the world. 
On Weibo, users post or repost content in a variety of domains. Since Weibo has a role similar to that of Twitter or Facebook in the United States, which are the main sources of data in some related works~\citep{pnas16, science18, science19, sa19, nc19}, a \textit{partially} aligned comparison for Weibo and U.S. platforms can be performed.

\subsection{Data Collection}\label{data:collection}
Almost all previous studies of Weibo data for empirical analysis ~\citep{liu-diffuse, epj-data} and detection~\citep{majing16, multimodal} used false news data collected from Weibo Community Management Center\footnote{\url{https://service.account.weibo.com/}} (hereafter, Center), an official platform to deal with user-reported violations of Weibo regulations. Reported posts that contain false information are fact-checked and made public by the platform. However, two biases determine which post is reported:
\begin{itemize}
	\item \textbf{Exposure bias.} Posts from influential users (e.g., celebrities) are exposed more frequently, enhancing collective wisdom for finding inaccuracy. In contrast, posts from little-known users with similar contents may not be noticed and reported as false. 
	\item \textbf{Selection bias.} The Weibo platform began operation in August 2009, but the reporting system started in 2012---the first false post was reported on May 29, 2012.\footnote{\url{https://service.account.weibo.com/show?rid=K1CaJ6wpc6aYf}} The lack of false news during the first three years may reduce the confidence of our results to reflect the overall situation. Moreover, we observe that users usually report posts related to their interests or reputations.
The last concern is that Weibo does not accept the reports of media accounts' publishing false information~\citep{weibo-convention}, so the Center data ignore false posts from media accounts. 
\end{itemize}

Thus, false news with little influence, with no clear involving user, or published by media accounts may not be well covered in the Center data. To adjust for these biases, we extended the dataset by tracing back from debunking posts or web articles from fact-checking sources (as Twitter researchers did~\citep{science18, liar, fakenewsnet}), including Zhuoyaoji,\footnote{\url{https://weibo.com/u/6590980486}} Weibo Piyao,\footnote{\url{https://weibo.com/weibopiyao}} Liuyanbaike,\footnote{\url{http://www.liuyanbaike.com/}} Jiaozhen,\footnote{\url{https://fact.qq.com/}} China Joint Internet Rumor-Busting Platform,\footnote{\url{http://www.piyao.org.cn/}} Jiangning Police Online,\footnote{\url{https://weibo.com/njjnga}} CCTV News,\footnote{\url{https://weibo.com/cctvxinwen}} and others. The process introduces multiple sources committed to debunking in different domains to neutralize the selection bias. And the search-and-sift step preserves any false news posts, including those popular and less popular, to tackle the exposure bias. The process was as follows:

We crawled data from the two aforementioned sources. For the Center data, we crawled the accessible false news posts from judging webpages.\footnote{\url{http://service.account.weibo.com/index?type=5&status=4}} For the data tracing from debunking information, we automatically extracted the query terms with our designed rules (e.g., extract text in quotation marks). For those not matching with the rules, the authors manually extract the query terms. Next, we searched the selected terms on the Weibo Search Engine\footnote{\url{https://s.weibo.com/}} and manually sifted out debunked false news from returning result lists. We double-checked each false news post and dropped those mistakenly sifted debunking posts and those posts with mismatching judging evidence. We crawled contents, publication date, and user profiles (specifically, gender, age, and verifying status) of each original post and each repost.
While we do not guarantee the thoroughness of data collected from the two sources above, the high coverage and richness of the dataset are the best it can be, under the limitations of data access. We end up with 44,728 false news posts, of which 24,690 are from the Center and 20,038 from the Weibo Search Engine, ranging from August 2009 to August 2019.

\subsection{Domain Annotation}
\textbf{Acquisition.} To assign a domain tag to each post in the dataset, we worked out the domain list and the classification criteria.\footnote{We collected existing category lists from aforementioned fact-checking websites and reports or papers on false news, including \textit{Zhuoyaoji}, \textit{Liuyanbaike}, \textit{Jiaozhen}, \textit{China Joint Internet Rumor-Busting Platform}, \textit{Tencent Rumor Governance Report}, and \textit{The spread of true and false news online} \citep{science18}. See details in S.1 of the supplementary material.}  Considering the the appropriateness of granularity and the congruence to the collected data, we identified nine domains: (1) Society \& Life, (2) Disasters \& Accidents, (3) Health \& Medicine, (4) Education \& Examinations, (5) Culture \& Sports \& Entertainment, (6) Science \& Technology, (7) Finance \& Business, (8) Politics, (9) Military. We may use the \textit{first} word of each domain as its short name for better representation. In the following analysis, the first five domains are classified as daily-life-related -- that is, events that have an impact on a person's daily life -- while the others are daily-life-unrelated (hereafter life-related and life-unrelated).

\textbf{Human Annotation.}
Because most debunking posts on Weibo lacked domain tags, we gathered 26 human annotators (graduate students) to code all posts into nine domains.
Following the existing researches (e.g., \citep{science18}) and the aforementioned fact-checking sites, we assigned only one domain tag for each post according to its main content. For those related to multiple tags, we labeled them by their key elements of interest. Consider the post: A Chinese military singer on active service named \textit{Dawei Jiang} has been naturalized in the United States.\footnote{\url{https://weibo.com/1603208240/zfGJMdgY2}. Debunked by the Center at \url{https://service.account.weibo.com/show?rid=K1CaJ6ANe66gf}} The post could be multi-labeled as Dawei Jiang is related to both military (a member of Chinese Army) and entertainment (a famous singer). However, his identity as a member of Chinese Army is the focus in this event, so the post was included in the domain Military in our dataset, not Entertainment.

In our workflow for annotation, the first author carefully labeled a randomly sampled subset of data containing posts in all domains (about 500 posts). Before the formal annotation, all annotators participated in a pilot annotation test. We showed 100 posts (selected from the first-author-annotated subset) and asked the annotators to assign domain labels. The annotators would go to the formal annotation with a hit rate higher than 0.8. As there could be different false news posts that were related to the same event (e.g., an earthquake), we first ran a K-means clustering based on TF-IDF vectors on the whole dataset. Then we split the annotation batches in each cluster to (ideally) let news posts in one batch be more likely to be in the same domain. An annotator could scan the batch in the same annotation page to speed up. The Cohen's Kappa coefficient is 0.76, which indicates good agreement. The first author carefully annotated the remaining posts, including those that were skipped during the annotation and those raising inter-annotator disagreement. We also randomly checked the posts in the same batches as the skipped or disagreement posts to improve the data quality.

\subsection{Data Overview}\label{data:overview}
The Weibo dataset consists of 44,728 false news posts, 24,690 from the Center and 20,038 from the Weibo Search Engine, ranging from August 2009 (when Weibo started operation) to August 2019. The posts were published by 40,215 users and reposted \textasciitilde3.4 million times. For each post, the contents, publication date, repost lists, and user profiles are attached. Note that Weibo allows a user to repost an original post multiple times with or without comments or replying text, so the repost lists are analogous to but not identical to those on Twitter.

\begin{table}[htbp]
\caption{Numbers and percentages of false news posts (cascades) in the Weibo dataset.}
\label{table:amount}
\centering
\begin{tabular}{|l|r|r|r|r|r|r|r|r|r|}
\hline
& Society & Health & Disasters & Culture & Education & Finance & Politics & Science & Military \\ \hline
\# & 16,003 & 7,788 & 4,515 & 3,866 & 3,527 & 3,402 & 2,560 & 2,302 & 765\\ \hline
\% & 35.78 & 17.41 & 10.09 & 8.64 & 7.89 & 7.61 & 5.72 & 5.15 & 1.71 \\  \hline
\end{tabular}
\end{table}

Table~\ref{table:amount} shows the domain-level distribution of false news posts on Weibo. The life-related domains are dominant domains on Weibo (79.8\%), which is in line with the finding that many false posts on Weibo are related to people's general concerns (most are life-related)~\citep{mis-in-china}. Political false news accounts for only 5.7\% on Weibo. This is quite different from the Twitter data in the U.S.~\citep{science18}, where politics is with the largest amount and more posts are from life-unrelated domains such as business and science. These differences may bring us statistical findings different from existing ones based on the Western data, which will be explored in the following sections.

\section{Diffusion of False News}\label{diffusion}
To answer \textbf{RQ1)}, we need to define and calculate domain-level capacity for diffusion. Before that, we introduce how to represent a cascade and measure its capacity for diffusion. Then we evaluate the domains by aggregating cascade-level scores.

\begin{figure}[h]
  \centering
  \includegraphics[width=0.6\linewidth]{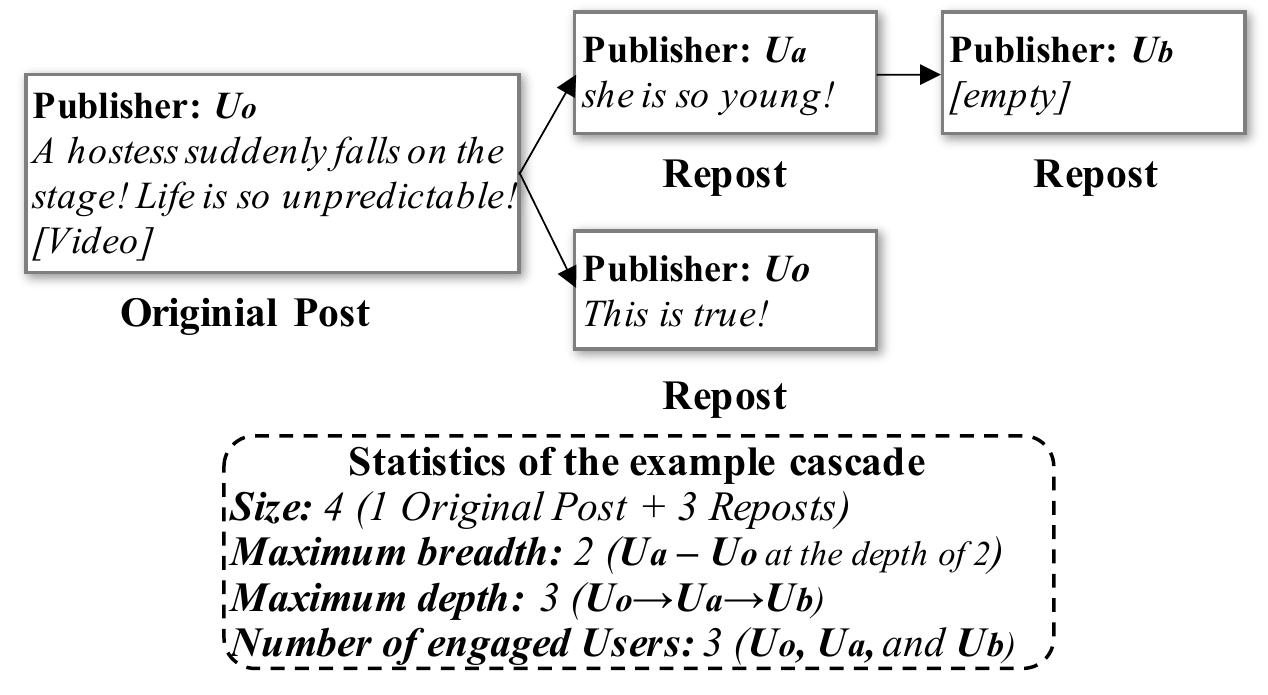}
  \caption{Example of a cascade and the statistics.}
  \label{fig:cascade}
\end{figure}

\subsection{Representing and Measuring Cascade-level Diffusion}
The spread of a news story on social media forms a \textit{cascade} with the original post (published by the original user, i.e., \textit{starter}) and reposts connected by reposting. However,  raw data does not provide tree-structure cascades. We exemplify a cascade with Figure~\ref{fig:cascade}. If a Weibo user $U_a$ reposts an original post from $U_o$ that reported a hostess's sudden death on a live show and said: ``She is so young!'' this repost will be displayed as \textit{$U_a$: she is so young!} (by default, no original post follows). If $U_b$ reposts this repost and said nothing, the repost of repost will be \textit{$U_b$: //$@U_a$: she is so young!} We pre-process the double-slashes format into  a tree structure by string split.

As the cascades are tree-structured, key attributes of trees could straightforwardly serve as the measurement of the diffusion of false news posts. Here, we use the widely used indicators to characterize tree structure, that is, size, maximum depth, and maximum breadth, to respectively indicate how many times that Weibo users participated in, how fierce the discussion was and how many engagements were individually triggered. Further, we consider the indicator, number of engaged users, to see the number of \textit{unique} users. Here are the definitions and illustrations:
\begin{itemize}
	\item \textbf{Size:} The number of posts in a cascade. Because there is only one original post in a cascade, the size equals the number of reposts plus one. In Figure~\ref{fig:cascade}, $U_o$'s original post is reposted by $U_a$, $U_b$, and itself, so the size of this cascade is 4.
	\item \textbf{Maximum depth:} The number of posts on the longest reposting path from the original one in a cascade. In practice, we recorded depths for each repost in a cascade. Thus the maximum of those recorded values was exactly the maximum depth. The maximum depth in Figure~\ref{fig:cascade} is 3, i.e., $U_o\rightarrow U_a \rightarrow U_b$.
	\item \textbf{Maximum breadth:} The maximum number of posts at any depth in a cascade. In practice, we obtain the frequency of each depth in a cascade and use the maximum of the frequencies as the maximum breadth. Since there is two reposts at the depth of 2, the maximum breadth in Figure~\ref{fig:cascade} is 2.
  \item \textbf{Number of engaged users:} The number of users engaging in a cascade, i.e., those having published at least a (re)post in the cascade. This indicator will be 3 in Figure~\ref{fig:cascade} as $U_o$, $U_a$, and $U_b$ have engaged in it.
\end{itemize}
 
\begin{figure}[t]
  \centering 
  \subfigure[]{\label{fig:ccdfs-a}
  \includegraphics[width=0.45\linewidth]{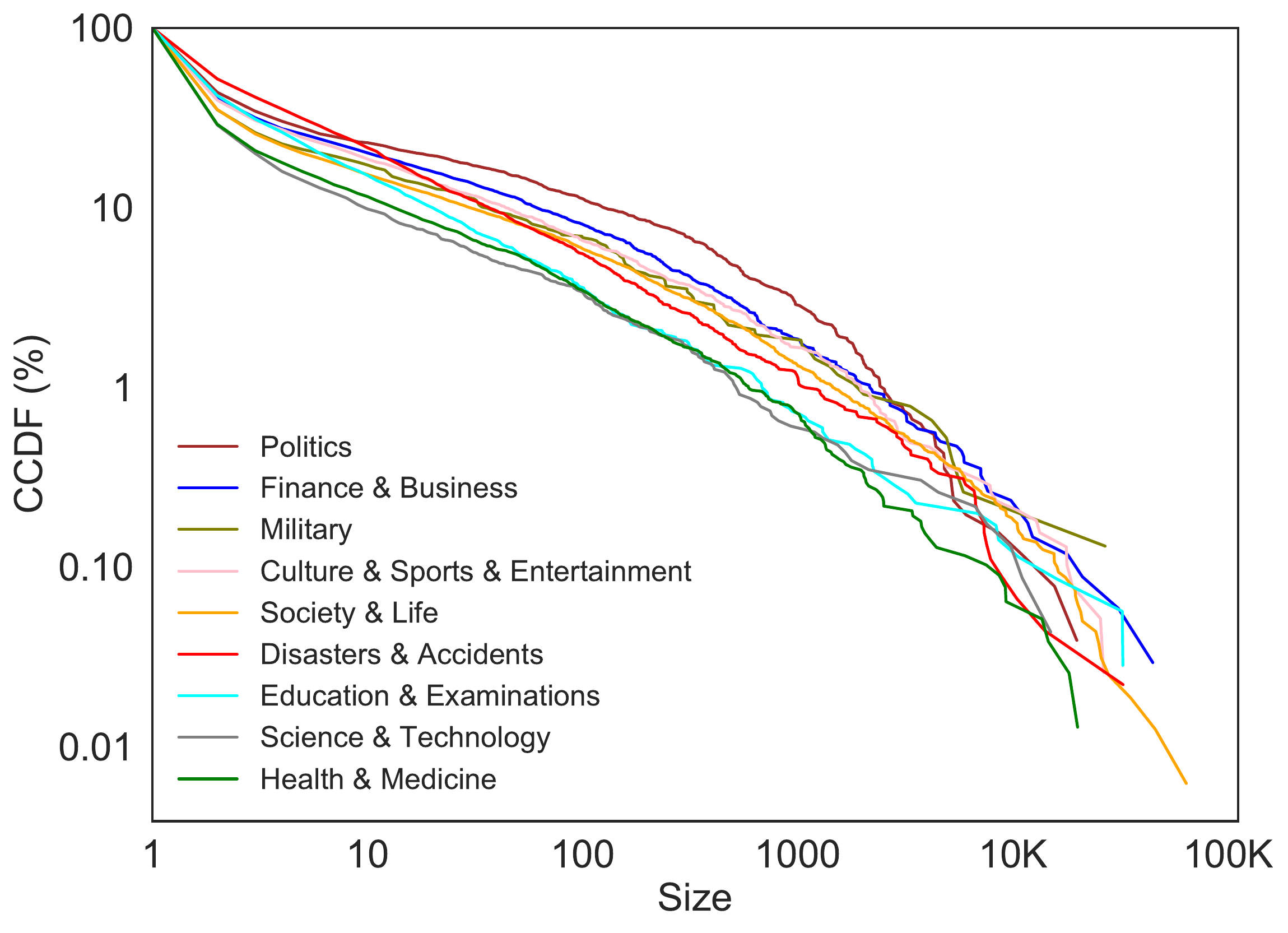}}
  \hspace{0.01\linewidth}
  \subfigure[]{\label{fig:ccdfs-b}
  \includegraphics[width=0.45\linewidth]{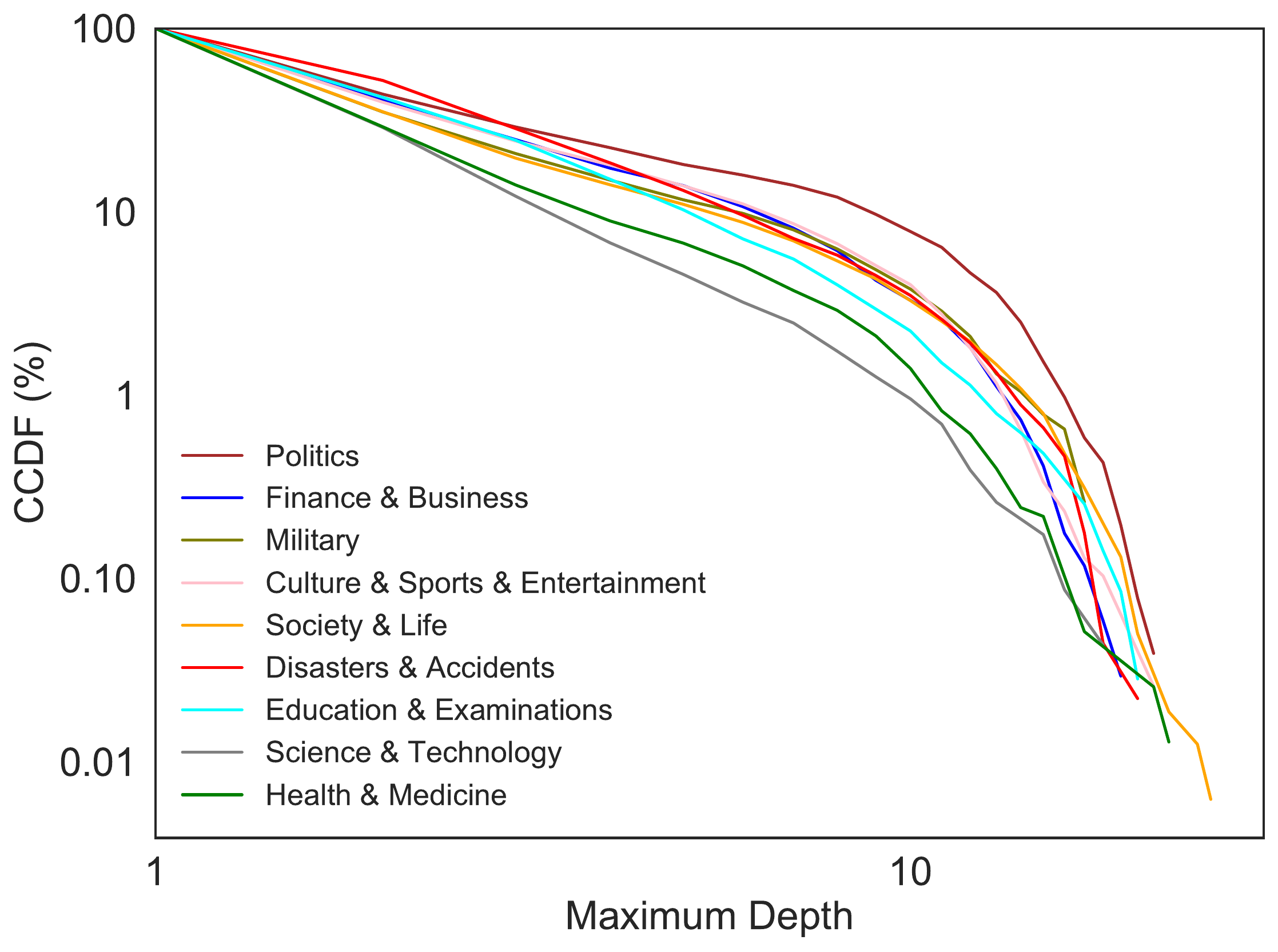}}
  \vfill
  \subfigure[]{\label{fig:ccdfs-c}
  \includegraphics[width=0.45\linewidth]{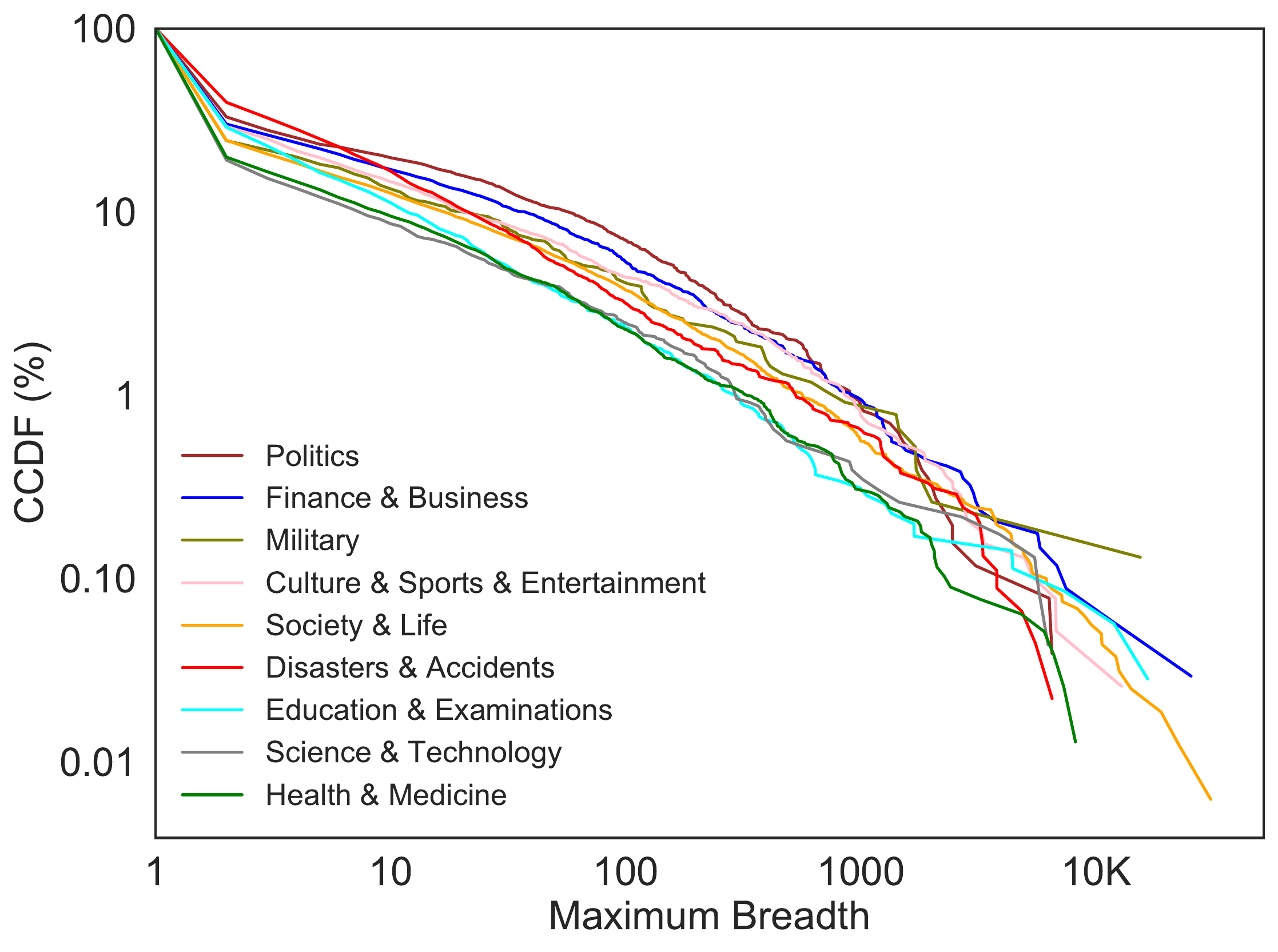}}
  \hspace{0.01\linewidth}
  \subfigure[]{\label{fig:ccdfs-d}
  \includegraphics[width=0.45\linewidth]{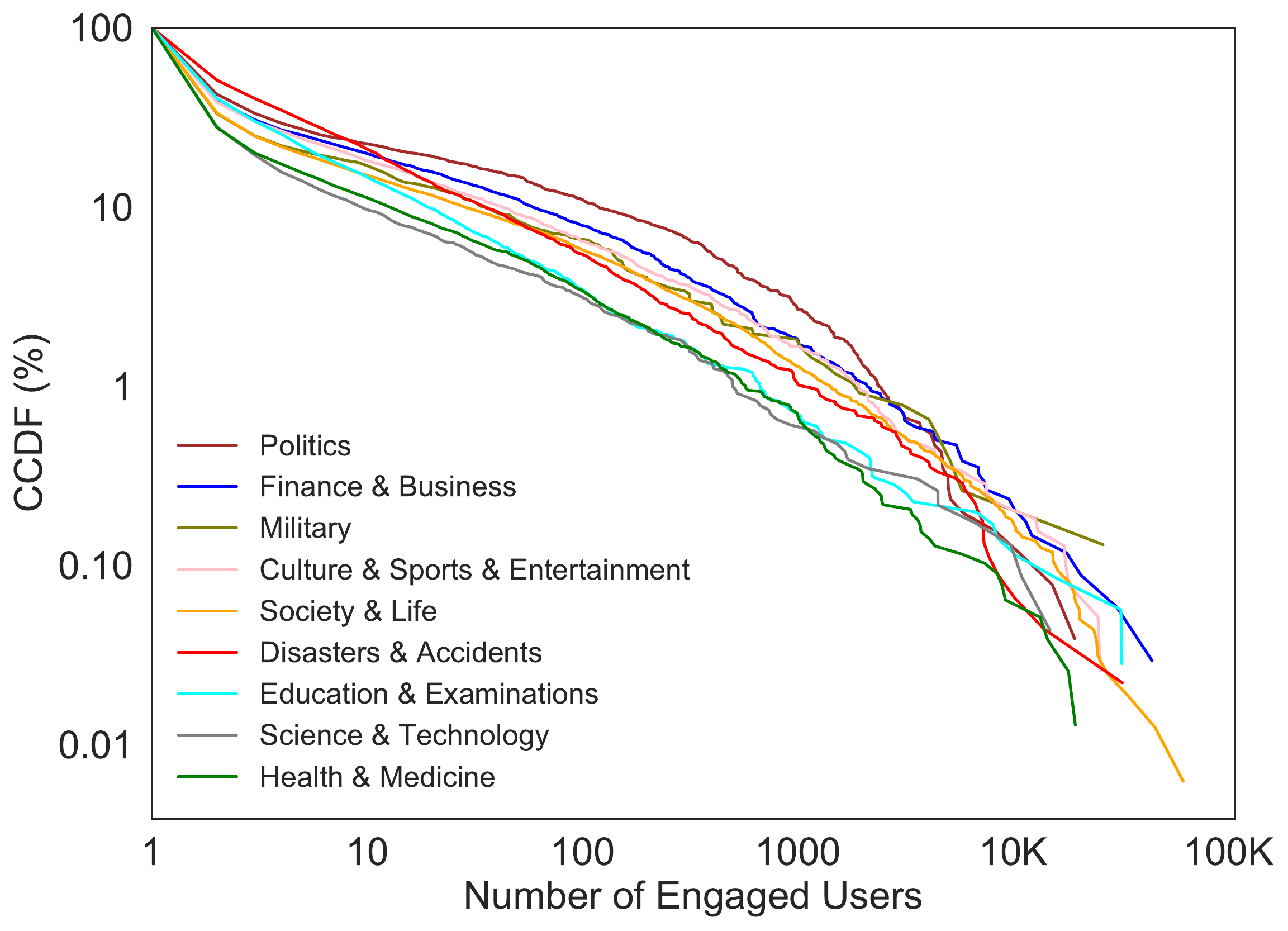}}
  \caption{Complementary cumulative distribution functions (CCDFs) of false news in nine domains. (a) Size. (b) Maximum Depth. (c) Maximum Breadth. (d) Number of Engaged Users. Both x-axis and y-axis are in logarithmic scales. Best viewed in color.}
  \label{fig:ccdfs}
\end{figure}

\subsection{Domain-level Capacity for Diffusion}\label{diffusion-capacity}
We aggregated the cascade-level results to obtain domain-level capacity for diffusion. For each domain, we first drew the Complementary Cumulative Distribution Function (CCDF) of cascades represented in the four indicators, as shown in Figure~\ref{fig:ccdfs}. Next, we calculated the areas under CCDF and added the areas (which were normalized within the nine areas of each indicator) up. The summation of normalized areas scores domain-level capacity for diffusion, as shown in Table~\ref{table:diff-rank}.

\begin{table}[htbp]
\small
\caption{Normalized areas under curve (NA) of the nine domains according to the corresponding measures and rankings (R) of the NAs. The column \textit{All} shows the summation of the four measures' NAs on the left columns and the corresponding rankings. The last column indicates the rankings of capacity for diffusion.}
\label{table:diff-rank}
\begin{tabular}{|l|r|r|r|r|r|r|r|r|r|r|}
\hline
\multirow{2}{*}{\textbf{Domain}} & \multicolumn{2}{r|}{\textbf{Size}} & \multicolumn{2}{r|}{\textbf{Max Depth}} & \multicolumn{2}{r|}{\textbf{Max Breadth}} & \multicolumn{2}{r|}{\textbf{\#Engaged Users}} & \multicolumn{2}{r|}{\textbf{All}} \\ \cline{2-11} 
 & \textbf{NA} & \textbf{R} & \textbf{NA} & \textbf{R} & \textbf{NA} & \textbf{R} & \textbf{NA} & \textbf{R} & \textbf{NA} & \textbf{R} \\ \hline
Politics & \textbf{1.000} & \textbf{1} & \textbf{1.000} & \textbf{1}  & 0.801 & 3 & 0.996 & 2 & \textbf{3.797} & \textbf{1}  \\ \hline
Finance & 0.985 & 2 & 0.764 & 4 & 0.996 & 2 & \textbf{1.000} & \textbf{1} & 3.745 & 2 \\ \hline
Military & 0.927 & 3 & 0.714 & 5 & \textbf{1.000} & \textbf{1} & 0.928 & 3 & 3.569 & 3 \\ \hline
Culture & 0.810 & 4 & 0.774 & 3 & 0.754 & 4 & 0.815 & 4 & 3.153 & 4 \\ \hline
Society & 0.737 & 5 & 0.688 & 7 & 0.697 & 5 & 0.741 & 5 & 2.863 & 5 \\ \hline
Disasters & 0.563 & 6 & 0.824 & 2 & 0.525 & 6 & 0.569 & 6 & 2.481 & 6 \\ \hline
Education & 0.488 & 7 & 0.694 & 6 & 0.481 & 7 & 0.494 & 7 & 2.157 & 7 \\ \hline
Science & 0.356 & 8 & 0.467 & 9 & 0.405 & 8 & 0.353 & 8 & 1.581 & 8 \\ \hline
Health & 0.307 & 9 & 0.519 & 8 & 0.332 & 9 & 0.311 & 9 & 1.469 & 9 \\ \hline
\end{tabular}
\end{table}
 
With the scores, we obtained the rankings for capacity for diffusion: Politics first, followed by Finance \& Business, Military, Culture \& Sports \& Entertainment, Society \& Life, Disasters \& Accidents, Education \& Examinations, Science \& Technology, and Health \& Medicine. In contrast to the domain-level cascade quantity distribution in Table~\ref{table:a2}, with one exception (false news in Science \& Technology), false news in life-unrelated domains has a more effective capacity for diffusion than that for life-related ones. That is, some life-related false posts were not as influential as those in life-unrelated domains.
 
Notably, of the nine domains, cascades on political false news are the largest (Figure~\ref{fig:ccdfs-a}), and deepest (Figure~\ref{fig:ccdfs-b}), the second-largest in terms of number of users reached (Figure~\ref{fig:ccdfs-c}) and the third in maximum breadth (Figure~\ref{fig:ccdfs-d}). In other words, despite the difference in the quantity of political false news on Weibo and Twitter~\citep{science18}, the capacity for diffusion of political false news is highly similar.

\section{Role of Engaged Users}\label{users}
Engaging with a story requires users to have a personal interest and to react to some immediate feelings about the story. It also, perforce, involves interaction with other users. To answer \textbf{RQ2}, \textbf{RQ3}, and \textbf{RQ4}, we explored these user effects based on user characteristics, emotions, and behaviors.

\subsection{User Characteristics}

We focus on three basic and accessible user attributes: gender, age, and account type.

\begin{figure}[h]
  \centering
  \includegraphics[width=0.8\linewidth]{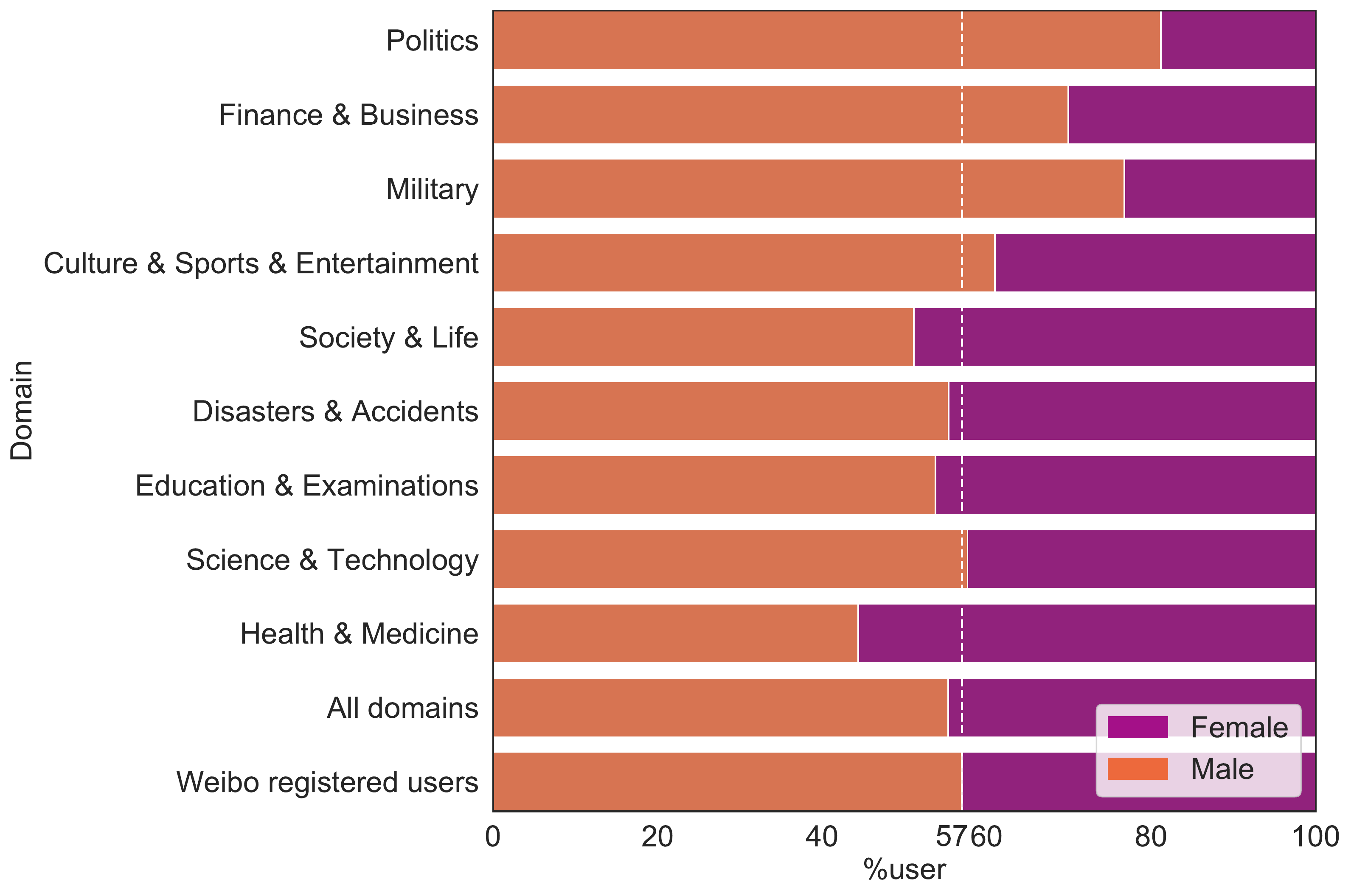}
  \caption{Gender Distribution. \textit{All domains} means the ratio across all nine domains and \textit{Weibo registered users} means the latest officially provided user gender ratio ~\citep{weibo2018report}. The white dash line at 57\%, is the proportion of male users on Weibo reported in~\citep{weibo2018report}. Best viewed in color.}
  \label{fig:gender}
\end{figure}

\paragraph{Gender} Figure~\ref{fig:gender} shows the gender distribution of false-news starters for each of the nine major domains. The gender ratio for all users on Weibo is ${\rm male:female}=57:43$~\citep{weibo2018report}. This is shown by the vertical white line in Figure~\ref{fig:gender} and in the last row of Figure~\ref{fig:gender}. Female and male users published about the same standardized percentage of false news posts, but we see a strong gender difference in news interests: Male users spread more false news on Politics, Military, and Finance \& Business, while female users spread more false news on Health \& Medicine, Society \& Life and Education \& Examinations. These gender differences in false-news interests are consistent with the diffusion and quantity measurements: The three male-dominated domains have more effective capacity for diffusion than the others, while the three female-dominated domains comprise larger amounts than others.

\begin{figure}[h]
  \centering
  \includegraphics[width=0.8\linewidth]{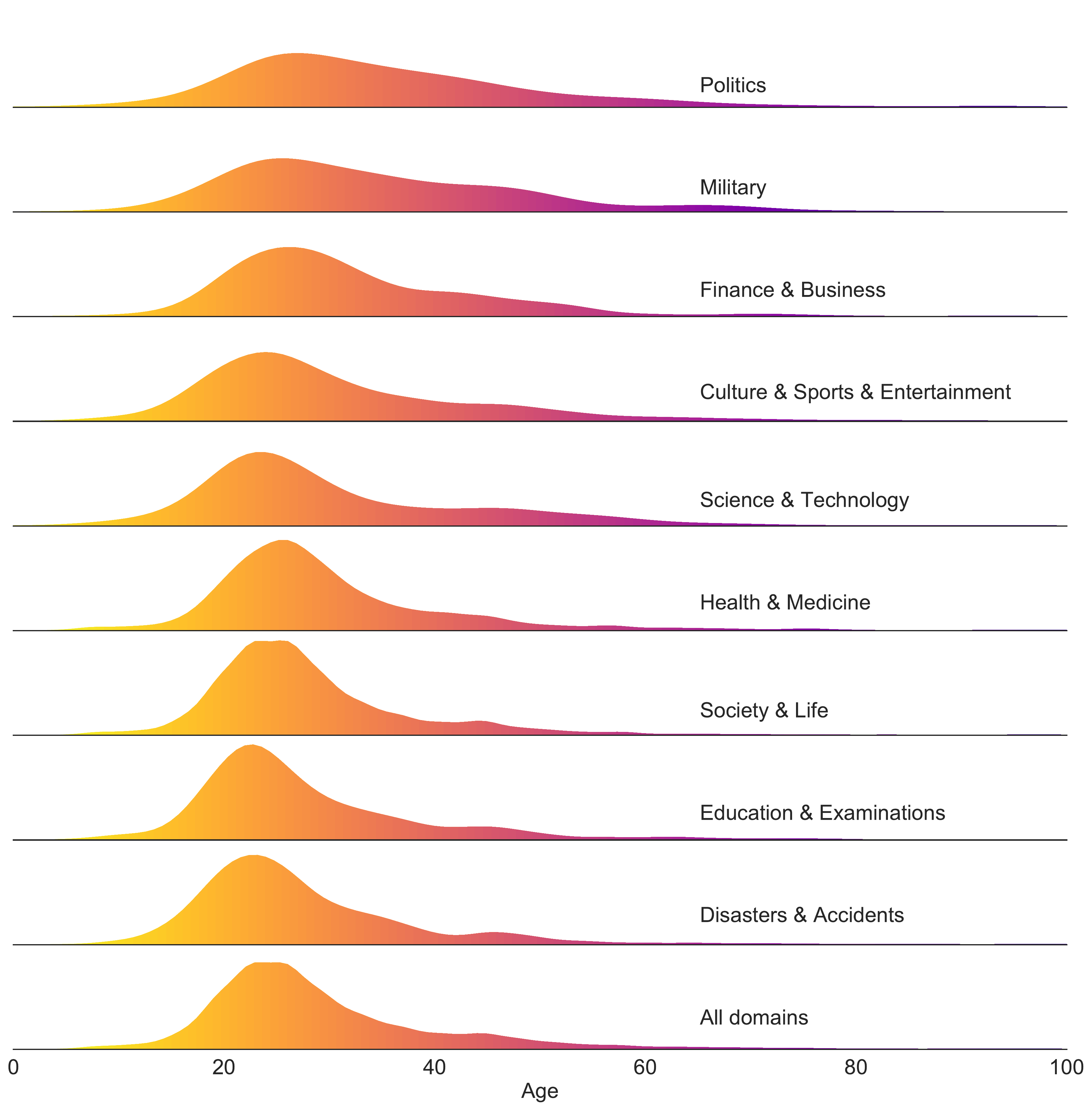}
  \caption{Age distribution in descending order according to the average age from top to bottom except \textit{All domains}. \textit{All domains} means the ratio across all nine domains. The age of a publisher used here was at the publication year. Best viewed in color.}
  \label{fig:age}
\end{figure}

\paragraph{Age} To filter out users with unreliable ages, we simply ignored posts from users who claimed to be under 6 or over 100 years old. We also excluded the verified organizational users. Figure~\ref{fig:age} shows the age distribution of false-news starters for each of the nine major domains. During the 2016 U.S. presidential election campaign, users over 65 were more likely to share articles of political false news~\citep{sa19}. We label this a \textit{seniors-attracted} tendency. In the Weibo data, we observed a similar but less pronounced seniors-attracted tendency for posts on Politics and Military matters.

Table~\ref{table:age} summarizes the statistics of the age distribution. In \citep{sa19}, 52.5\% of political false news shares were associated with users over 65 and 7.9\% with users under 30, while on Weibo, the over-65 group published 2.6\% of political false news and users under 30 contributed 43.1\%. Across all the nine domains, 64.9\% of false news posts in our Weibo data were published by users under 30, which is lower than the proportion of users under 30 of the Weibo population (81\%) ~\citep{weibo2018report}.

\begin{table}[h]
\small
\caption{Age statistics of publishers of all domains. The last row shows Weibo's overall data in its official report~\citep{weibo2018report}.}
\label{table:age}
\begin{tabular}{|p{0.2\textwidth}|r|r|r|r|r|}
\hline
\textbf{Domain} & \textbf{Aver. Age} & \textbf{Upper Quartile} & \textbf{$<$30} & \textbf{30$\sim$65} & \textbf{$>$65} \\ \hline
Politics & 34.56 & 42 & 337(43.0\%) & 425(54.3\%) & 21(2.7\%) \\ \hline
Finance & 31.93 & 38 & 513(51.9\%) & 457(46.2\%) & 19(1.9\%) \\ \hline
Military & 34.17 & 42 & 113(44.8\%) & 129(51.2\%) & 10(4.0\%) \\ \hline
Culture  & 30.43 & 36 & 810(59.2\%) & 530(38.7\%) & 29(2.1\%) \\ \hline
Society & 27.98 & 31 & 4,767(69.3\%) & 2,054(29.9\%) & 53(0.8\%) \\ \hline
Disasters  & 27.14 & 31 & 1,362(71.0\%) & 540(28.2\%) & 16(0.8\%) \\ \hline
Education  & 27.65 & 32 & 944(70.0\%) & 390(28.9\%) & 15(1.1\%) \\ \hline
Science & 30.29 & 37 & 498(62.3\%) & 291(36.4\%) & 11(1.4\%) \\ \hline
Health & 29.50 & 33 & 2,049(63.3\%) & 1,142(35.3\%) & 46(1.4\%) \\ \hline
All domains & 29.04 & 33 & 11,393(64.8\%) & 5,958(33.9\%) & 220(1.2\%) \\ \hline
Weibo~\citep{weibo2018report} & In 23$\sim$30 group & In 23$\sim$30 group & $\sim$81\% & \multicolumn{2}{r|}{$\sim$19\%} \\ \hline
\end{tabular}
\end{table}

\paragraph{Account Type} Each account is either a verified or unverified account. Verified accounts include verified individuals and verified organization. Weibo ensures that a verified user's identity on its profile is authentic, generally increasing the user's credibility and influence.\footnote{\url{https://verified.weibo.com/}}
Table~\ref{table:account-type} shows proportions of the three types of accounts as cascade starters and reposts they led to in each domain. Unverified users published 73.1\% of all false news, provoking 42.1\% of all reposts, while verified users published 26.9\% of false news posts, provoking 57.9\% of reposts.
This largely challenges the intuitive or commonsense interpretation that verified users, which less than 0.1\% of Weibo users were verified~\citep{weibo2015report}, are more credible.
Although, as described in Section~\ref{data}, we did our best to collect less popular false posts (which were more possibly published by unverified users with fewer followers) as Section~\ref{data:collection} described, the verified users spread false news posts over its account type proportion. This corresponds to the finding on Twitter~\citep{science18}. In the verified users, individual users published more false posts than organizational users across all domains except Science \& Technology. In a sense, verified individuals, who were influential (that is, generally had more followers than unverified ones) but lacked professional information source and rigorous editorial process, contributed the most to the spread of false news. 
 
\begin{table}[htbp]
\small
\caption{Distribution of different type of accounts as the cascade starters (\%st) and reposts they led to (\%rp) in the nine domains.}
\label{table:account-type}
\begin{tabular}{|p{0.18\textwidth}|r|r|r|r|r|r|}
\hline
\multirow{2}{*}{\textbf{Domain}} & \multicolumn{2}{r|}{\textbf{Unverified User}} & \multicolumn{2}{r|}{\textbf{Verified Individual}} & \multicolumn{2}{r|}{\textbf{Verified Organization}} \\ \cline{2-7} 
 & \textbf{\%st} & \textbf{\%rp} & \textbf{\%st} & \textbf{\%rp} & \textbf{\%st} & \textbf{\%rp} \\ \hline
Politics & 75.70 & 58.80 & 18.32 & 35.39 & 5.98 & 5.81 \\ \hline
Finance & 57.94 & 31.90 & 22.22 & 46.07 & 19.84 & 22.03 \\ \hline
Military & 75.82 & 65.89 & 16.86 & 31.16 & 7.32 & 2.95 \\ \hline
Culture & 69.19 & 37.53 & 19.56 & 56.76 & 11.25 & 5.72 \\ \hline
Society & 76.60 & 42.80 & 14.51 & 49.06 & 8.89 & 8.14 \\ \hline
Disasters  & 75.11 & 39.51 & 20.07 & 34.78 & 4.83 & 25.71 \\ \hline
Education  & 71.31 & 42.83 & 15.03 & 49.39 & 13.67 & 7.78 \\ \hline
Science  & 66.25 & 45.52 & 16.16 & 38.14 & 17.59 & 16.34 \\ \hline
Health & 74.92 & 35.47 & 13.97 & 47.99 & 11.11 & 16.55 \\ \hline
All Domains & 73.08 & 42.10 & 16.38 & 46.35 & 10.53 & 11.55 \\ \hline
\end{tabular}
\end{table}

In contrast, while verified organizational users generally have advantages in their access to information sources and have more incentives to control the quality of published content, they still published more than 10\% of false news posts. To evaluate the organizational users' ability to distinguish false stories, we analyzed their belief in what they reposted\footnote{We did not observe the original posts here because they could only provide evidence on users' inability to tell truth from falsehood, not the opposite.}. Specifically, we focus on the role of users who represent six verified organizations: police, government (gov.), media, schools, companies (biz., excluding media run as companies), and social organizations (social org.).\footnote{\textit{Police} here refers to the government-run public security departments. We separated \textit{Police} from \textit{Government} because one of its functions is to defend cyberspace security~\citep{mps-functions}, including combating fake news (e.g., punish false rumor spreaders after Tianjin blast~\citep{mps-actions}).} These organizations are generally credible in the eyes of Chinese citizens especially those state-run~\citep{government-dividend}, so posts labeled as coming from these organizations are followed by many users. We evaluated the content in a repost by an organizational user as an indicator of whether \textit{it} believed the false news story being passed on. If no disbelief is expressed, we infer that the user believes the story and is motivated to help spread it. In contrast, if the added content expresses disbelief or doubt, then we infer that the user reposts as a way of mitigating the potential misconceptions of others regarding the veracity of the story.

We classified the reposts of the engaged organizational users into five classes: \textit{believe}, \textit{debunk}, \textit{do not believe} (for short, \textit{DNB}), \textit{doubt}, and \textit{unknown} (i.e., neutral or unrelated content). A repost with no added content is labeled as \textit{believe} by default, except when the user reposts a debunking repost. As Table~\ref{table:account-orgs} shows, 85.0\% of all reposts from users associated with one of the six organizations ($n=18,057$) showed belief in reposts of false news, with a range of 70.9\% for users associated with the media to 88.4\% for users associated with companies. Summing \textit{debunk}, \textit{DNB}, and \textit{doubt}, the media had by far the highest rate of disbelief (26.0\%), close to the disbelief rate (23.1\%) found on Twitter for news organizations~\citep{liquanzhi16}. By comparing the proportions of the nine domains' posts in the fooling-organization list and all false (see Table~\ref{table:account-diff}), we found that organizational users were slightly more likely to be fooled by false news on Politics, Finance \& Business, and Society \& Life.

\begin{table}[htbp]
\small
\caption{Proportion of reposts in the five classes for each type of verified organizational users. The lowest proportion in Believe and the highest in other classes are in \textbf{boldface.}}
\label{table:account-orgs}
\begin{tabular}{|l|r|r|r|r|r|r|r|}
\hline
 & \textbf{Police} & \textbf{Gov.} & \textbf{Media} & \textbf{Biz.} & \textbf{School} & \textbf{Social Org.} & \textbf{Total} \\ \hline
\textbf{Believe} & 0.769 & 0.855 & \textbf{0.709} & 0.884 & 0.878 & 0.804 & 0.850 \\ \hline
\textbf{Debunk} & 0.074 & 0.032 & \textbf{0.093} & 0.005 & 0.015 & 0.007 & 0.023 \\ \hline
\textbf{DNB} & \textbf{0.143} & 0.099 & 0.121 & 0.048 & 0.068 & 0.096 & 0.069 \\ \hline
\textbf{Doubt} & 0.005 & 0.004 & \textbf{0.046} & 0.032 & 0.034 & 0.037 & 0.030 \\ \hline
\textbf{Unknown} & 0.009 & 0.010 & 0.031 & 0.030 & 0.005 & \textbf{0.056} & 0.027 \\ \hline
\end{tabular}
\end{table}

\begin{table}[htbp]
\small
\caption{Number and proportion of false news posts that fooled organizational users and those were reposted in the nine domains. The last column shows the differences and positive values are in \textbf{boldface}.}
\label{table:account-diff}
\begin{tabular}{|l|r|r|r|}
\hline
\textbf{Domain} & \textbf{\# (\%) of fooling-org original posts} & \textbf{\# (\%)of all reposted false original posts} & \textbf{Percentage Difference} \\ \hline
Politics & 263 (0.087) & 1,122 (0.067) & \textbf{0.019} \\ \hline
Finance & 310 (0.102) & 1,381 (0.083) & \textbf{0.019} \\ \hline
Military & 39 (0.013) & 268 (0.016) & -0.003 \\ \hline
Culture & 250 (0.082) & 1,531 (0.092) & -0.009 \\ \hline
Society & 1,102 (0.363) & 5,623 (0.337) & \textbf{0.026} \\ \hline
Disasters & 278 (0.092) & 2,350 (0.141) & -0.049 \\ \hline
Education & 258 (0.085) & 1,484 (0.089) & -0.004 \\ \hline
Science & 115 (0.038) & 664 (0.040) & -0.002 \\ \hline
Health & 418 (0.138) & 2,265 (0.136) & 0.002 \\ \hline
\end{tabular}
\end{table}

At 26.0\% disbelief, however, even the most skeptical of the users (those associated with the media) were highly likely to be taken in by false news posts. The result is not in line with the finding that journalists are more likely to deny false rumors on Twitter during three crisis events \citep{journalists-kate}. We argue that institutions, like the Police and the Media, which generally enjoy solid reputations among citizens for their authority~\citep{credibility-authority}, especially in China~\citep{government-dividend}, are not skillful at maintaining the reliability of what they repost. This may facilitate the diffusion of false news instead of containing it.

\subsection{User Emotions}\label{emotion}
Inspired by ~\citep{dual-emotion}, we evaluated the emotions in the publishers' contents and engaged users' responses respectively in each domain. We adopted the affective lexicon ontology database~\citep{emotion-dict} curated by Dalian University of Technology, China. This database comprises 27,467 Chinese words, each of which is manually classified into one of seven emotion types with intensity: joy, like, anger, sadness, fear, disgust, and surprise. This list represents a Chinese adaptation to the original list of six universal emotion types from Eckman~\citep{eckman72, eckman87}. It adds ``like'' to that list.\footnote{According to ~\citep{emotion-dict}, \textit{like} is to represent the fine-grained emotions including respect, praise, belief, and love.} For each (re)post, we first segmented the texts by using the Chinese lexical analyzer, THULAC~\citep{thulac}, and then recorded the intensity of the corresponding emotion if a word matched with an affective lexicon ontology from ~\citep{emotion-dict}. We used a seven-dimensional vector (denoted as ${\rm e}$) to record the intensities of each kind of emotion and then normalized each entry in the vector with the summation of intensities. For example, ${\rm e}=[0,0,0.2,0,0.5,0.3,0]$ means that the text expressed 20\% anger, 50\% fear, and 30\% disgust. We finally ranked the nine domains according to the average intensity in each emotion.

\begin{figure}[h]
  \centering
  \includegraphics[width=\linewidth]{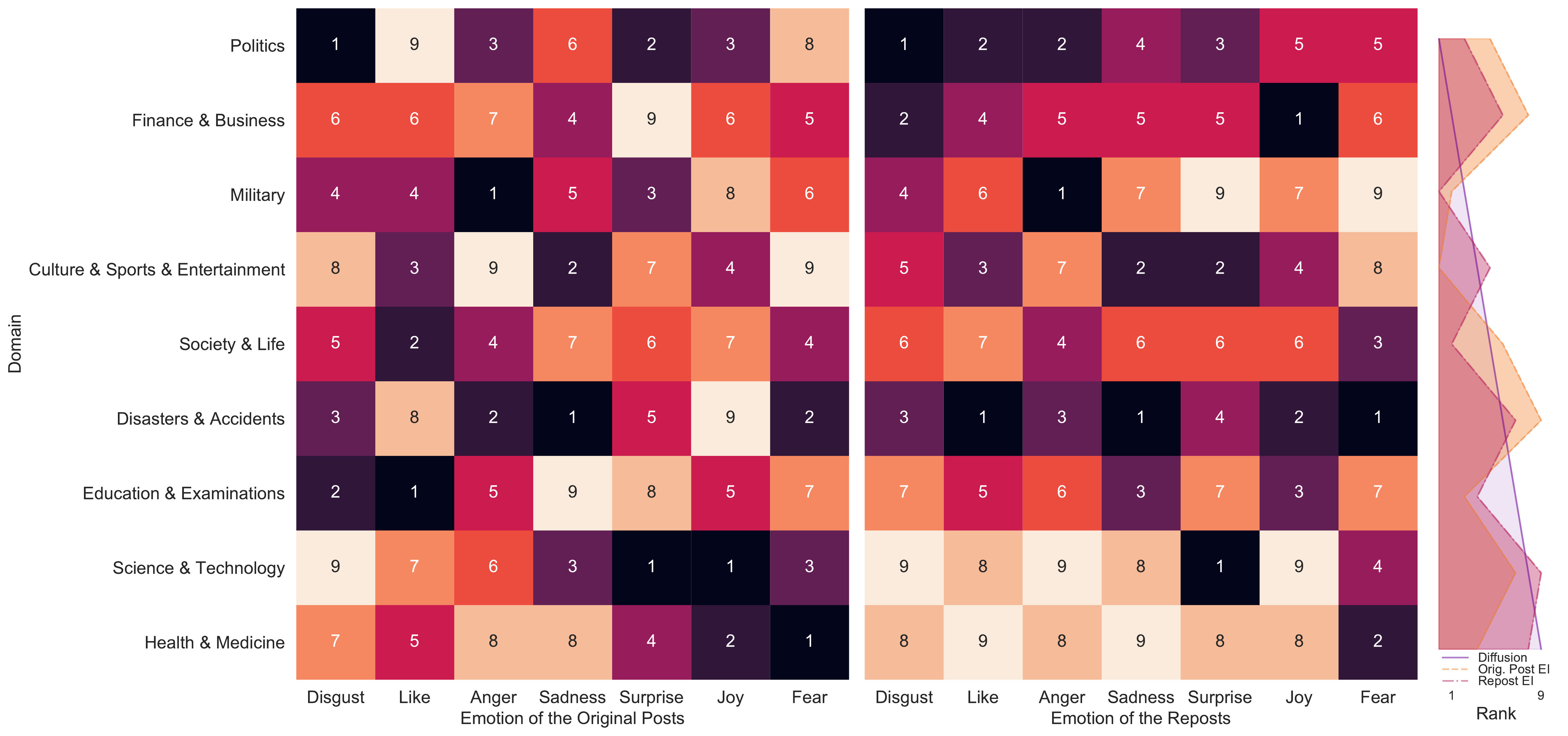}
  \caption{Domain-emotion heatmaps with intensity ranks in the original posts (left) and the reposts(middle). The rank numbers are in the grids. Right: The rank of the whole emotional intensities (EI) for each domain in the original posts (Orig. Post EI) and reposts (Repost EI). \textit{Diffusion} represents the rank of capacity for diffusion in Section~\ref{diffusion-capacity}. Best viewed in color.}
  \label{fig:emotion}
\end{figure}

The left and middle part of Figure~\ref{fig:emotion} show the domain-emotion heatmaps with the intensity ranks of the original posts and reposts, respectively. We obtained the whole ranking by averaging out the emotional intensities of all the original posts (or reposts) across domains and compared it with the ranking of capacity for diffusion (Section~\ref{diffusion-capacity}) in the right part of Figure~\ref{fig:emotion}. The whole emotional intensity of the reposts was more related to the capacity for diffusion ($\rho = 0.61, P=0.08$), compared to that of the original posts ($\rho = 0.05, P=0.90$). This indicates that while the original posts may not be that emotional, they are good at provoking the emotion of engaged reposting users. 

At the domain-level, false news on Politics and Finance \& Business showed weak emotion in the original posts, but inspired strong emotion in the reposts, while the opposite was the case for posts on Military, Science \& Technology, and Health \& Medicine. At the emotion-level, false news in the domains with more effective capacity for diffusion provoked more disgust, anger, and like in the reposts ($\rho=0.88, 0.70$, and $0.60$ respectively). \citet{science18} found that false news inspired responses expressing greater surprise and greater disgust on Twitter and \citet{anger-covid} found anger contributed to the spread of COVID-19 misinformation. Here we had the consistent finding on disgust and anger, but surprise was almost unrelated to the capacity for diffusion ($\rho = -0.13$ in the original posts, $0.10$ in the reposts). Please refer to Table~\ref{table:a2} and Table~\ref{table:a3} in~\ref{appendix-c} for details.

We conclude that, while false news items may provoke strong emotions like disgust and anger, and therefore, high item capacity for diffusion, the content of the original posts does not need to be emotional.

\subsection{User Behaviors}\label{behavior}
Besides statistical observation of reposting (Section~\ref{diffusion-capacity}), we focus here on how specific reposting behaviors \textit{promoted} the spread. In our data, we observed two special behaviors that could significantly increase the size of cascades and thus promote the spread: (1) Reposting in high frequency. One might repeatedly repost an original post to maximize its visibility in different time period and towards different user groups; and (2) Reposting as a reply (i.e., replying to an engaged user while reposting). One might argue with other users by replying again and again to a (re)post and thereby promote the spread of a story. To measure the two behaviors, we designed the following measurements:
\begin{itemize}
	\item \textbf{Cascade concentration score:} the proportion of reposts that is not a user's first engagement in one cascade. It evaluates the users' level of engagement. The equation is $${\rm c=\left(1-\frac{\#uniqueUsers}{\#reposts}\right)\times 100\%}$$where ${\rm c}$ is the score and \# is short for ``the number of''.
	\item \textbf{Number of replies:} the number of reposts that start with ``Reply to @user'' where \textit{user} represents the user to be replied to. It evaluates the level of interactions among the engaged users. 
\end{itemize}

\begin{table}[htbp]
\small
\caption{Cascade concentration score (CCS for short, \%), number of replies, and starter engagement in the nine domains and of the total. Only cascades with at least one repost were considered for averaging number of replies out. Std is short for standard error. The maximum values are in \textbf{boldface}.}
\label{table:behavior}
\centering
\begin{tabular}{|l|r|r|r|}
\hline
\multirow{2}{*}{\textbf{Domain}} & \multirow{2}{*}{\textbf{CCS}} & \multicolumn{2}{r|}{\textbf{Number of Replies}} \\ \cline{3-4} 
 &  & \textbf{Mean} & \textbf{Std} \\ \hline
Politics & \textbf{2.60} & 1.799 & 6.984 \\ \hline
Finance & 2.04 & 1.067 & 5.569\\ \hline
Military & 2.57 & \textbf{1.881} & 9.348 \\ \hline
Culture & 1.94 & 0.965 & 5.445 \\ \hline
Society & 2.00 & 0.847 & 5.032 \\ \hline
Disasters  & 2.04 & 0.458 & 2.847\\ \hline
Education  & 2.23 & 0.423 & 3.037 \\ \hline
Science  & 1.20 & 0.259 & 1.495 \\ \hline
Health  & 1.56 & 0.493 & 3.072\\ \hline
Total & 1.95 & 0.793 & 4.689  \\ \hline
\end{tabular}
\end{table}

Table~\ref{table:behavior} shows the cascade concentration score and number of replies in the nine domains. Reposts of false news in the domains where posts were generally widely diffused were concentrated and interactive. At 2.60, political false news had the highest concentration score among the nine domains. That is, to achieve the same number of reposts, political false news needed fewer engaged accounts on average than did that in other domains. There are 0.793 replies across all domains, but 1.881 on average for military false news in a single cascade, 2.4 times the overall average and 7.2 times the 0.259 figure for false news on Science \& Technology.

High concentrated and interactive spread of false news suggests there might be some users playing a special role in promoting the spread. Here we investigate the role of cascade starters (i.e., the users publishing the original posts) because publishers usually tend to attract more people and persuade the reposters to believe the content. Counting only cascades with at least one repost, political false news has the highest starter engagement of all nine domains. Fully 18.89\% of cascade starters reposted the original post in Politics and top 1.87\% starters reposted at least 10 times. Starters in other domains also engaged in cascades, but the engagements are not comparable to that for Politics (Table~\ref{table:starter}).

\begin{table}[htbp]
\small
\caption{Statistics on the starter engagement of cascades of nine domains and area under curve with rankings. Std is short for standard error. \textit{\% of having $\geq$ N reposts} means the percentage of cascades with over \textit{N} starters' reposts, \textit{NA} is the normalized area under CCDF curve, and \textit{R} is the ranking of \textit{NA} in each domain. The maximum values are in \textbf{boldface}.}
\label{table:starter}
\centering
\begin{tabular}{|l|r|r|r|r|r|r|r|}
\hline
\textbf{Domain} & \textbf{Mean} & \textbf{Std} & \textbf{Max} & \textbf{\% of having $\geq$ 1 repost} & \textbf{\% of having $\geq$ 10 reposts} & \textbf{NA} & \textbf{R} \\ \hline
Politics & \textbf{2.351} & 12.878 & \textbf{343} & \textbf{18.89} & \textbf{1.87} & \textbf{1.000} & \textbf{1} \\ \hline
Finance  & 1.345 & 2.269 & 73 & 14.26 & 0.43 & 0.445 & 6 \\ \hline
Military & 1.474 & 2.532 & 40 & 18.66 & 0.37 & 0.516 & 3 \\ \hline
Culture  & 1.234 & 1.318 & 35 & 12.34 & 0.20 & 0.371 & 8 \\ \hline
Society& 1.394 & 3.713 & 164 & 14.41 & 0.46 & 0.454 & 5 \\ \hline
Disasters& 1.233 & 1.284 & 37 & 11.32 & 0.30 & 0.370 & 9 \\ \hline
Education  & 1.373 & 3.657 & 121 & 14.15 & 0.34 & 0.456 & 4 \\ \hline
Science  & 1.286 & 2.224 & 44 & 8.58 & 0.45 & 0.413 & 7 \\ \hline
Health & 1.483 & 6.689 & 275 & 12.76 & 0.75 & 0.521 & 2 \\ \hline
\end{tabular}
\end{table}

Though starters' reposting essentially promoted the spread, there might be starters reposting for correcting false claims instead of convincing others. However, as far as we know, no method can be directly applied to understand starters' real motivation with only digitally collected social media data. Here, applying the results of the content analysis on the posts of organizations, if a starter reposts a false news story without expressing any disbelief, we take that as \textit{prima facie} evidence of the starter's motivation. For starters who reposted their original posts, we classified their reposts into four classes by merging the category of debunking into \textit{do not believe} (\textit{DNB}) because we could not expect the starters to debunk effectively. In 2,280 labeled cascades (5.1\% of the total, 10.7\% of those reposted), only 12.5\% of starters reposted at least once for expressing disbelief (\textit{DNB} and \textit{doubt}) regarding the content of the original posts, indicating that they might not post false news deliberately at the start of a cascade. The remaining starters, who always reposted with belief, tried to promote the spread of false news.

\begin{table}[htbp]
\small
\caption{Numbers and proportions of the cascades that starters expressed disbelief in the nine domains. The last row showed the whole statistics.}
\label{table:starter-disbelief}
\centering
\begin{tabular}{|l|p{0.26\textwidth}|p{0.2\textwidth}|p{0.12\textwidth}|}
\hline
\textbf{Domain} & \textbf{\# of cascades in which starters expressed disbeliefs} & \textbf{\# of cascades that starters reposted} & \textbf{Disbelief rate (\%)} \\ \hline
Politics & 32 & 212 & 15.09 \\ \hline
Finance  & 55 & 197 & 27.92 \\ \hline
Military & 9 & 50 & 18.00 \\ \hline
Culture & 69 & 189 & 36.51 \\ \hline
Society  & 216 & 810 & 26.67 \\ \hline
Disasters& 107 & 266 & 40.23 \\ \hline
Education  & 90 & 210 & \textbf{42.86} \\ \hline
Science  & 10 & 57 & 17.54 \\ \hline
Health  & 79 & 289 & 27.34 \\ \hline
All domains & 667 & 2,280 & 29.25 \\ \hline
\end{tabular}
\end{table}

\begin{table}[htbp]
\small
\caption{Examples of main types of starters' reposts without disbelief. The text in \textbf{boldface} represents the latest reposts from the starter at that moment. We replace the three attached pictures in the original post with text here and anonymize all usernames. The original post is omitted due to the space limit and the reposts are translated into English.}
\label{table:starter-example}
\begin{tabular}{|p{0.13\textwidth}|p{0.4\textwidth}|p{0.41\textwidth}|}
\hline
\textbf{Type} & \textbf{Example Cascade} & \textbf{Explanation} \\ \hline
thanking the users who share the same opinion & \textbf{Thanks for your comments!} // @UA: Reply to @Starter: As a national civil servant, he raped a woman by taking advantage of his position, causing her to be pregnant and give birth to a daughter. Want to get away with it? Take my advice: You'd better be responsible for the mother and daughter! // @Starter: Repost // (The original post) & User UA expressed agreement to the original post. As a reply, the starter reposted and thanked UA. \\ \hline
replying to the comment which challenges the story & \textbf{The girl took those photos, just in case!} // @UB: How could she take photos when being drunk? // @UC: // @UD: Repost // (The original post) & User UB questioned about how the attached photos, which recorded the rape, was taken. Then, the starter gave its (unconvincing) explanation.\\ \hline
using techniques of neutralization & \textbf{My attached photos are confirmed by Baidu! It's unnecessary to focus on them!} // (The original post) & The starter was reported publishing false news because the attached photos were from another story. Then the starter reposted and tried to shift the responsibility (to Baidu) and divert readers' attention. The starter used the technique of neutralization named denial of responsibility. \\ \hline
\end{tabular}
\end{table}

At the domain level, the disbelief rate (here, the proportion of starter-engaged cascades in which the starters show disbelief) for false news on Politics, Military, and Science \& Technology was low (from 15.09\% to 18.00\%), while it was high for Education \& Examinations, Disasters \& Accidents, and Culture \& Sports \& Entertainment had high disbelief rates (from 36.51\% to 42.86\%) (Table~\ref{table:starter-disbelief}). When a starter continues reposting the false news post published by itself with little disbelief, it may be promoting other users to discuss the story, thus achieving the goal of spreading. By observing 333 reposts without disbelief by the highly engaged starters, we found these starters often repost by thanking users who share the same opinion, replying to comments that challenge the story, or using techniques of neutralization like denial of responsibility~\citep{neutralization}. Examples are shown in Table~\ref{table:starter-example}. This provides an alternative hypothesis about why false news in domains that comprise a small proportion of the total, like Politics and Military, have effective capacity for diffusion: The stories may inspire the starters to actively promote the spread process.

\section{Discussion} \label{discussion}

Our research is to understand the role of \textit{domain} in the spread of false news. We performed our research in two steps: We first measured the capacity for diffusion of false news in each domain (\textbf{RQ1}), and then explored the related factors in user characteristics, emotions, and behaviors (\textbf{RQ2}, \textbf{RQ3}, and \textbf{RQ4}, respectively). In this section, we first answer the \textbf{RQ}s and analyze them based on our key findings in the context of existing studies. Next, we introduce how our findings can help improve practical systems on this issue and recommend several future research directions.

\subsection{Answering the Research Questions}

\textbf{RQ1) Are there differences in the capacity for diffusion of false news in different domains?}
The capacity for diffusion of false news varied from domain to domain. False news in life-unrelated domains diffused more effectively than that in life-related ones and political false news had the most effective capacity for diffusion. However, the ranking of capacity for diffusion and that of the amount are quite different: Life-related false posts are more than life-unrelated ones and Politics ranked the third to last, which is aligned with the finding that Chinese users did not perceive much political false news and concerned with false news relevant to their daily life or well-being~\citep{government-dividend}. 
The discrepancy can be explained in two ways. On the one hand, ordinary users hardly have the source of life-unrelated rumors beyond social media. Even if they do, they may be too cautious to publish them with concern about the possible punishments~\citep{government-dividend}. On the other hand, reposting is a behavior with much fewer consequences because the platform only punishes who publishes false stories (the root user of a cascade)~\citep{weibo-regulation}.

\textbf{RQ2) How are the demographic factors, like gender, age, and verification status, related to the spread of false news in different domains?}
We found a slight age and gender effect: Male, older users were more likely to publish false news in domains where posts were generally widely diffused, while female, younger users were more related to others.

As to the account type, the proportion of false news posts published by verified users and the reposts they led to largely exceeded its proportion in all users. Verified organizational users mostly reposted false information with belief, indicating their inability to recognize the falsehood. Dangers arose since verified users, individual or organizational, lacked fact-checking ability that matched their influence.
Generally, an organizational account is managed by human teams (mostly employees in the public relationship or marketing departments). In this sense, the ``inability'' may be caused by two factors: the employee's carelessness and the absence of an internal editorial process. A recent case confirmed our assumption to some extent: A governmental employee mistakenly used the official account of the local earthquake agency to repost an entertainer-related post because the employee forgot to change to the personal account~\citep{earthquake-agency}. Without a necessary internal editorial process before (re)posting a message through a verified organizational account, these accounts will inevitably face a risk of encountering reputation crises.

\textbf{RQ3) How are the emotional signals related to the spread of false news among the domains?}
We found that the emotional signals in user responses (reposts) were more related to capacity for diffusion of false news than those in contents (original posts). On the one hand, with high-arousal emotions, users tend to repost or comment, leading to the posts go viral~\citep{content-viral, content-viral-anger}. However, on the other hand, it is unnecessary to use emotional language to arouse readers' strong emotions. For example, false stories related to controversial objects or persons could provoke emotional responses even with no emotional words included.

\textbf{RQ4) What did the engaged users do that promoted the spread of false news in each domain?}
User engagements were more concentrated and interactive in false news cascades in domains where posts were generally well-diffused. In the engaged users, false news starters were more proactive in interacting with other users. Most starters' reposts were to attract more people to read or convince those skeptical engaged users, not to debunk and mitigate the spread, especially in the domains where posts were generally widely diffused. This finding bridges a connection between false news spread and controversy arising---the behaviors of engaged users promote the spread of false news by making the discussion more controversial.

\subsection{Connection and Comparison with Existing Studies}
The findings in this study confirm or contradict the conclusions of existing studies, which are summarized as follows:

1) We derived a \textit{similar} diffusion capacity ranking (life-unrelated ``$>$'' life-related) on Chinese Weibo data as that on English Twitter data~\citep{science18}, though the distribution of false news posts is quite \textit{different} (e.g., Twitter $>$ Weibo in terms of \%Politics). This validates that false news in domains such as Politics and Finance is consistently more likely to incite readers to repost, regardless of the context of languages and perception levels.

2) We provide new evidence that male, older users were more likely to publish political false news, as \citet{science19} did before the 2016 U.S. presidential election, based on a much longer period without targeting a specific political event. However, this trend did \textit{not} hold true in the picture of nine domains. We found that female, younger users contributed more to the domains such as Health, Society, and Education. This indicates that the connection between false news publishing and user gender or age is not constant but depends on domains.

3) We observed a counter-intuitive phenomenon that verified users were actually vulnerable to false news. Our results showed that the verified users did fail in veracity judgment of social media posts with a high probability and thus, were not as credible as expected.
This warns us to reevaluate the role of verified users in false news detection~\citep{user-profile2, user-profile3, GCAN}.

4) We found that the strong emotions in reposts rather than original posts were more related to false news spread. Unlike existing empirical works that observed either original posts~\citep{moral-emotion} or reposts~\citep{science18} for differentiating true and false news, we relate them with the false news spread across domains for comparison. By comparing with existing findings, we found that the emotions such as anger and disgust might constantly serve as a motivating factor for the spread of false news.

5) We highlighted the property of well-diffused false news cascades: concentrated, interactive, and starter-proactive. Unlike existing network-based methods which focus on community property (e.g.,~\citep{jinzhiwei-aaai16,network-zhouxinyi}), we provide a new observation of reposting behavior itself.

\subsection{Implications on System Design}
Typically, a false news detection system~\citep{zhouxing2015, defend-system, quin} has the following procedures:

\begin{itemize}
	\item \textbf{Suspicious news discovery:} Collects news posts that are suspicious as candidates~\citep{rumor-resolution, check-worthy}. Often formulated as a ranking task~\citep{checkthat2021}.
	\item \textbf{News veracity prediction:} Use fake news detection methods (mostly from multiple perspectives) to predict news veracity.
	\item \textbf{Display and explanation:} Show the abnormal elements (e.g., propagation network, questions from the comments, and contradiction with known facts) to explain to users why the news might be false.
\end{itemize}

Our findings indicate the necessity of dividing and conquering false news in different domains in the detection system. The suggestions are as follows:

\textbf{1) For suspicious news discovery, prioritize news from domains where posts are generally well-diffused.} In practice, the step to ``find out'' candidate news often takes more resources than the veracity prediction; thus, a careful design of the initial filtering strategy is important for maintaining good scalability and efficiency. Unfortunately, very few works provide guidance. We reveal that life-unrelated false news has a more effective capacity for diffusion than life-related. Thus, a system with limited computing resources should prioritize suspicious posts in domains such as Politics, Military, and Finance \& Business, or set a higher frequency of fetching news in these domains.

\textbf{2) For news veracity prediction, integrate models' outputs with awareness of domains.} The existing system tends to use a single model or the integration of a model set for \textit{all} news posts. Our findings on the differences among domains suggest the challenge of ``one solution for all domains.'' For example, many health-related false news posts had weak diffusion, and thus few engaged users, so the propagation-based or user-based models may fail to capture useful signals and judge them incorrectly. In this sense, a wise solution is to fact-check against external knowledge bases~\citep{health-misinformation}. In contrast, political false news posts may spread widely but the truth might be unknown for now. Prediction based on propagation networks would be more practical.

\textbf{3) For display and explanation, benchmark the properties with a domain-level statistic.} To enrich the result page, existing systems often list the properties of the given post such as the attributes of engaged users. We argue that this is not informative and helpful to let the audience know the effects of the numbers. Given that the user effects were quite different among domains, we suggest that these properties should be benchmarked with the statistics of all historical posts in the same domain. For example, adding a note like ``over $x$\% of false news in [domain]'' below the number of how many verified users engaged in. 

\subsection{Contributions to Future Research}
\paragraph{Highlighting the role of domains in false news research}
Our findings on the spread and user effects of false news in nine domains uncovered that there existed common and unique features for false news in different domains. It indicates that some findings in a specific domain or event may not generalize to other domains and those on mixed-domain data may ignore unique characteristics of less popular domains. Therefore, to have a clear picture of false news, considering the role of domains in future research is highly recommended.

\paragraph{Providing a partial solution to infer the beliefs and motivations with limited data}
We analyzed the comments in reposts of organizational users and starters to infer their beliefs (and belief shifts if applicable) to false news posts. The results unveiled the organizational users' lacking ability to recognize false news and the starters' most willingness to promote the spread. Though our comment-based inference is only effective when the textual or behavioral signals exist, it provides the way to infer users' minds and motivations when no more psychological information is provided.

\paragraph{Constructing a new Chinese false news dataset} We analyzed the shortcoming of existing Weibo false news datasets and proposed that retrieving false news posts that scattered in the platform was important to mitigate selection and exposure biases of data collection. In this way, we collected a new Weibo false news dataset containing false events excluded by the Center data. Our bias-mitigated collecting method can be a reference for future works.

\subsection{Limitations and Future Work} \label{limitations}

\paragraph{Mixed Factors} Along with the report on the false news on Weibo, we have compared some of our findings with existing ones on Twitter and Facebook. Although we found an interesting phenomenon that political false news on both Weibo and Twitter had the most effective capacity for diffusion, attracted similar user groups in terms of age and gender, and provoked similar emotions, differences existed in other aspects such as amounts and emotions. Our comparison suggested that although there were more relevant studies, the findings based on the U.S. social media data was not general enough to be a global proxy. Indeed, multiple variables such as country (China vs. the U.S.), language (Chinese vs. English), and platform (Weibo vs. Twitter/Facebook) led to the differences in several ways. However, the respective effects of sociocultural backgrounds, language use, and platform managements were hard to measure from these data. To know the influence of these factors, in-lab experiments with variable control or data from more diverse platforms are needed.

\paragraph{Research methods} We combined statistical analysis and content understanding to obtain meaningful results, some of which were strongly related to users' internal states, such as starters' and organizational users' beliefs to false news. Our analysis may be limited by the existence of explicit textual signals and annotators' understanding of texts. Considering that conducting a user survey or interview after, what for original users would be 12 years of activity, is unfeasible, follow-up surveys or in-the-moment interviews (e.g., just after a user is told of having published a false news post) are potential methods to further observe the phenomena reported in this retrospective study.

\section{Conclusions} \label{conclusion}
We performed analysis on multi-domain false news on Weibo from 2009 to 2019. On Weibo, political false news, though few, has the most effective capacity for diffusion. Broadly, life-unrelated domains have more effective capacity for diffusion than life-related domains on Weibo, though the number of the latter exceeds that of the former (79.8\% vs. 20.2\% respectively). Our observations of user effects show that a widely diffused false news post on Weibo is associated strongly with certain types of users (male, old, or verified users), provokes strong emotions in the repost list, and evokes more replies in a limited group with the starter's promotion. However, the gender and age effects were mostly due to the news consumption preferences of gender and age groups.

Based on our findings, we highlight the roles that the domain plays in practical false news detection systems. We made suggestions on the pipeline design including suspicious news discovery, veracity prediction, and proper display.
Our findings also point to issues for further research on false news in China and other countries: First, in addition to political false news, we need more research in other domains, because some findings in Politics may not apply to others. Second, we advocate for more focus on users who have special roles, like starters and verified users. Third, false news on social media is a global issue. Comparing false news on the U.S. and Chinese platforms is a start, but it is clear from our analysis that no single platform can serve as a global template for understanding and further mitigating false news. More work on diverse platforms will help us determine features of false news that are common across countries and languages and the unique ones as well. This, in turn, will help us all in facing the challenge of false news effectively.

\section*{Acknowledgments}
The authors thank Carole Bernard, Xirong Li, and Amrita Bhattacharjee for their proofreading and feedback on the manuscript. The research work is supported by the National Key Research and Development Program of China (2021AAA0140203), and the Zhejiang Provincial Key Research and Development Program of China (2021C01164). 

\appendix

\section{Domain-level average emotion intensities}\label{appendix-c}

\begin{table}[htbp]
\caption{Average intensity of each domain in seven types of emotions by measuring the original posts. Spearman coefficients between the emotions of the original posts and capacity for diffusion in the nine domains are: 0.40 (Disgust), -0.23 (Like), 0.30 (Anger), 0.20 (Sadness), -0.13 (Surprise), -0.33 (Joy), and -0.65 (Fear).}
\label{table:a2}
\centering
\begin{tabular}{|p{0.12\textwidth}|p{0.1\textwidth}|p{0.08\textwidth}|p{0.08\textwidth}|p{0.1\textwidth}|p{0.11\textwidth}|p{0.08\textwidth}|p{0.08\textwidth}|}
\hline
\textbf{Domain} & \textbf{Disgust} & \textbf{Like} & \textbf{Anger} & \textbf{Sadness} & \textbf{Surprise} & \textbf{Joy} & \textbf{Fear} \\ \hline
Politics & 0.3100 & 0.3336 & 0.0102 & 0.0489 & 0.0188 & 0.0980 & 0.0253 \\ \hline
Finance  & 0.1852 & 0.4231 & 0.0060 & 0.0627 & 0.0051 & 0.0857 & 0.0496 \\ \hline
Military & 0.1965 & 0.4656 & 0.0193 & 0.0537 & 0.0131 & 0.0747 & 0.0463 \\ \hline
Culture & 0.1595 & 0.4880 & 0.0026 & 0.0996 & 0.0073 & 0.0976 & 0.0223 \\ \hline
Society  & 0.1899 & 0.5012 & 0.0095 & 0.0480 & 0.0078 & 0.0835 & 0.0527 \\ \hline
Disasters  & 0.2176 & 0.3540 & 0.0174 & 0.1078 & 0.0085 & 0.0691 & 0.0905 \\ \hline
Education  & 0.2398 & 0.5261 & 0.0088 & 0.0299 & 0.0060 & 0.0883 & 0.0285 \\ \hline
Science & 0.1242 & 0.3726 & 0.0076 & 0.0975 & 0.0322 & 0.1419 & 0.0818 \\ \hline
Health  & 0.1756 & 0.4281 & 0.0035 & 0.0459 & 0.0094 & 0.1115 & 0.1245 \\ \hline
\end{tabular}
\end{table}

\begin{table}[htbp]
\caption{Average intensity of each domain in seven types of emotions by measuring the reposts. Spearman coefficients between the emotions of the reposts and capacity for diffusion in the nine domains are: 0.88 (Disgust), 0.60 (Like), 0.70 (Anger), 0.33 (Sadness), 0.10 (Surprise), 0.43(Joy), and -0.48 (Fear)}
\label{table:a3}
\begin{tabular}{|p{0.12\textwidth}|p{0.1\textwidth}|p{0.08\textwidth}|p{0.08\textwidth}|p{0.1\textwidth}|p{0.11\textwidth}|p{0.08\textwidth}|p{0.08\textwidth}|}
\hline
\textbf{Domain} & \textbf{Disgust} & \textbf{Like} & \textbf{Anger} & \textbf{Sadness} & \textbf{Surprise} & \textbf{Joy} & \textbf{Fear} \\ \hline
Politics & 0.3100 & 0.3336 & 0.0102 & 0.0489 & 0.0188 & 0.0980 & 0.0253 \\ \hline
Finance  & 0.1852 & 0.4231 & 0.0060 & 0.0627 & 0.0051 & 0.0857 & 0.0496 \\ \hline
Military & 0.1965 & 0.4656 & 0.0193 & 0.0537 & 0.0131 & 0.0747 & 0.0463 \\ \hline
Culture & 0.1595 & 0.4880 & 0.0026 & 0.0996 & 0.0073 & 0.0976 & 0.0223 \\ \hline
Society & 0.1899 & 0.5012 & 0.0095 & 0.0480 & 0.0078 & 0.0835 & 0.0527 \\ \hline
Disasters  & 0.2176 & 0.3540 & 0.0174 & 0.1078 & 0.0085 & 0.0691 & 0.0905 \\ \hline
Education  & 0.2398 & 0.5261 & 0.0088 & 0.0299 & 0.0060 & 0.0883 & 0.0285 \\ \hline
Science  & 0.1242 & 0.3726 & 0.0076 & 0.0975 & 0.0322 & 0.1419 & 0.0818 \\ \hline
Health & 0.1756 & 0.4281 & 0.0035 & 0.0459 & 0.0094 & 0.1115 & 0.1245 \\ \hline
\end{tabular}
\end{table}

\printcredits

\bibliographystyle{cas-model2-names}

\bibliography{cas-refs.bib}

\end{document}